\documentclass[pra,aps,groupedaddress,twocolumn,floatfix]{revtex4}

\usepackage[T1]{fontenc}
\usepackage[english]{babel}
\usepackage[dvips]{graphicx}
\usepackage{enumerate,amsthm,amsmath,amssymb,color,graphicx,bbm,float}
\usepackage{natbib}
\usepackage{hyperref}
\newtheorem{theo}{Theorem} 
\newtheorem{lemma}[theo]{Lemma}

\begin{document}

\title{A symmetrization technique for continuous-variable quantum key distribution}

\author{Anthony Leverrier}
\affiliation{ICFO-Institut de Ciencies Fotoniques, Mediterranean Technology Park,
08860 Castelldefels (Barcelona), Spain}

\date{\today}

\begin{abstract}
We introduce a symmetrization technique which can be used as an extra step in some continuous-variable quantum key distribution protocols. By randomizing the data in phase space, one can dramatically simplify the security analysis of the protocols, in particular in the case of collective attacks. The main application of this procedure concerns protocols with postselection, for which security was established only against Gaussian attacks until now. Here, we prove that under some experimentally verifiable conditions, Gaussian attacks are optimal among all collective attacks.
\end{abstract}

\maketitle

Quantum key distribution (QKD) is the art of distilling a secret key among distant parties, Alice and Bob, in an untrusted environment. The remarkable feature of QKD is that it is secure in an information theoretic sense \cite{SBC08}. QKD protocols come in two flavors depending on the type of quantum measurement they use: either a photon counting measurement for \emph{discrete-variable} protocols or a homodyne detection for \emph{continuous-variable} (CV) QKD. While the security of the former is now rather well understood (with the notable exceptions of the differential phase shift \cite{IWY02} and coherent one-way \cite{SBG05} protocols), security of CV protocols has been more elusive (see \cite{WPG11} for a recent review). This is mainly due to the fact that the infinite dimensional Hilbert space required to describe these protocols makes the analysis quite challenging. 

Among all CVQKD schemes, the protocol GG02 is certainly the easiest one to analyze \cite{GG02}. In this protocol, Alice sends $n$ coherent states $|\alpha_k\rangle = |x_{2k} + i x_{2k+1}\rangle$ to Bob who measures the states he receives either with a homodyne detection (thus randomly choosing one quadrature to measure for each state) or a heterodyne detection (in which case, Bob measures both quadratures at the same time). Alice's modulation is Gaussian, meaning that $x_k$ is a centered normal random variable with a given variance. In the case of a heterodyne detection for instance \cite{WLB04}, Bob obtains a classical vector ${\bf y} = [y_1,\cdots, y_{2n}]$ which is correlated to Alice's vector ${\bf x}= [x_1,\cdots, x_{2n}]$. In this paper, we use bold font to refer to vectors. Then, using parameter estimation, reconciliation and privacy amplification, they can extract a secret key. For this specific protocol, Gaussian attacks (where the action of the eavesdropper can be modeled by a Gaussian quantum channel between Alice and Bob) are known to be optimal among collective attacks \cite{GC06,NGA06,PBL08}. Using de Finetti theorem and conditioned upon an extra verification step \cite{RC09}, these collective attacks are actually optimal in general in the asymptotic limit. The only step which is currently missing in this security analysis is a tight reduction from coherent to collective attacks in the finite-size regime \cite{LGG10}.

The security status of other CVQKD protocols is far less advanced. In particular, not much is known for protocols using postselection \cite{SRL02,LKL04,LSS05}. In these protocols, the idea is that Alice and Bob will only use some data to distill the secret key and discard the rest. More precisely, they only keep the data compatible with a positive key rate. This method, inspired by advantage distillation techniques, certainly makes the protocol more robust against imperfections such as losses or noise in the channel, and potentially gives the best practical CVQKD protocol (see Refs. \cite{LSS05,LRH06} for experimental implementations). It is therefore of considerable importance to be able to assess its security. Unfortunately, there are currently no full security proof for this scheme, not even against collective attacks. 
In fact, the only result available so far is an analysis in the case of Gaussian attacks \cite{HL07,SAA07}.
This is, however, far from being sufficient for two reasons: first, Gaussian attacks are not believed to be optimal against this protocol; second, one can never prove in practice that a given quantum channel is indeed Gaussian. 
The problem is that the only tool currently available to establish the security of a CV protocol against collective attacks, namely Gaussian optimality \cite{WGC06} does not seem to help much for protocols with postselection (see, however, a recent approach along those lines in Ref. \cite{WSR11}).

In this paper, we introduce a new proof technique based on a symmetrization procedure that allows us to make some progress concerning the security analysis of CVQKD with postselection. In particular, we will show how this symmetrization allows us, under some verifiable conditions, to consider that the quantum channel is indeed Gaussian, even though the physical channel may actually be non-Gaussian. This then means that checking the security against Gaussian attacks (which can be done with present tools) is indeed sufficient to get full security against collective attacks, and in fact against arbitrary attacks in the asymptotic limit thanks to de Finetti theorem \cite{RC09}.

\section{A symmetrized protocol}
The usual technique to prove the security of a Prepare and Measure (PM) protocol such as GG02 where Alice sends coherent states to Bob who measures them, is to consider an equivalent Entanglement-Based (EB) protocol. In the latter, Alice prepares two-mode squeezed vacuum states of which she measures one mode with a heterodyne detection and sends the second mode to Bob. Interestingly, before Alice and Bob measure their respective modes, their share a bipartite state $\rho_{AB}$. In this paper, we will restrict our attention to collective attacks, meaning that at the end of the protocol, Alice and Bob share $n$ copies of that state, that is, $\rho_{AB}^{\otimes n}$. 

In general, one cannot perform a perfect tomography of this state, simply because it lives in an infinite dimensional Hilbert space. In the case of protocols without postselection, this is not a problem since the secret key rate can be safely computed from the the Gaussian state with the same first two moments as $\rho_{AB}$  \cite{GC06,NGA06}. This is remarkable because one only needs to compute the covariance matrix of the state instead of its whole density matrix. 

Unfortunately, this approach fails in the case of protocols with postselection. Indeed, one would then need to compute the covariance matrix of the state \emph{given it was postselected}. In principle, one could do this analysis with the experimental data obtained from the EB version of the protocol; but one cannot directly reconstruct this covariance matrix from the data observed in the actual PM version of the protocol. 
Indeed, the probabilistic map corresponding to a successful postselection is too complicated and one cannot expect to analyze its effect on general non-Gaussian states. For this reason, our only hope for a security proof seems to be to somehow enforce the Gaussianity of the state $\rho_{AB}$ Alice and Bob will use in their protocol. The idea is therefore to add an extra step to the usual protocol, that will make the state $\rho_{AB}^{\otimes n}$ more Gaussian. Let us note $\mathcal{S}$ the quantum map induced by this symmetrization.

It is now clear that if one had $\mathcal{S}\left(\rho_{AB}^{\otimes n}\right) = \rho_G^{\otimes n}$ where $\rho_G$ is the bipartite Gaussian state with the same covariance matrix as $\rho_{AB}$, then the security of the symmetrized protocol against general collective attacks would be identical to the security of the original protocol against Gaussian attacks. One could then compute the secret key rate, simply from the transmission and excess noise of the quantum channel, exactly as in the case of protocols without postselection. The symmetrization we introduce below will not induce an exact Gaussification, but only an approximate one. However, the quality of the Gaussification, characterized by the fidelity between  $\mathcal{S}\left(\rho_{AB}^{\otimes n}\right)$  and $\rho_G^{\otimes n}$ will increase with $n$, and tend to 1 if some experimentally verifiable conditions (on the moment of order 4 of $\rho_{AB}$) are met. 

The symmetrization we consider here was introduced in \cite{LKG09} where it was argued that it corresponds to the natural symmetry for protocols using a Gaussian modulation of coherent states. 
In the EB scenario, before they both perform their heterodyne measurements, Alice and Bob would apply random conjugate passive linear transformations over their $n$ modes. Once this is done, they apply the usual postselection protocol.
This symmetrization can also be used in the PM scenario, and crucially, one can simply apply it to the \emph{classical} data ${\bf x}$ and ${\bf y}$ of Alice and Bob. 
More concretely, in the PM scenario, Alice and Bob follow the standard scenario of sending coherent states and performing a heterodyne detection for Bob. Then, Alice draws a random transformation with the Haar measure on the group $K(n):=O(2n,\mathbb{R}) \cap Sp(2n,\mathbb{R})$ (isomorphic to the unitary group $U(n)$), that is the transformations corresponding to linear passive transformations in phase-space. She informs Bob of her choice of transformation (over the authenticated classical channel), and both parties apply this transformation to their respective $2n$-vectors ${\bf x}$ and ${\bf y}$.

\section{Equivalence between the EB and PM symmetrized protocols}
In order to study the security of the symmetrized PM protocol, one needs to show that its equivalent EB protocol corresponds to the one symmetrized through the application of random conjugate passive transformations in phase-space. It is useful to introduce three different distributions that can be used to describe the two scenarios. In order to simplify the exposition, let us first consider the analysis of a generic CVQKD protocol in the case of collective attacks, meaning that the protocol is entirely described by a single use of the quantum channel. First, the PM protocol is naturally described by a joint probability distribution $P(x_1, x_{2}, y_1, y_{2})$ where $x_1,x_2$ (resp. $y_1,y_2$) refers to Alice's (resp. Bob's) measurement results. The EB scenario is characterized by the bipartite state $\rho$ shared by Alice and Bob before their respective measurements. In the context of a CV protocol, it is natural to describe this state by its Wigner function $W(x_1,x_2,y_1,y_2)$ where index 1 (resp. 2) refers to the first (resp. second) quadrature of Alice or Bob's mode. 
Alternatively, since we restrict ourselves to protocols where both Alice and Bob perform a heterodyne detection, we can also consider the convenient characterization in terms of the  $Q$-function $Q(x_1,x_2,y_1,y_2)$  of the state $\rho$. This $Q$-function corresponds to the probability distribution sampled from when measuring the state with heterodyne detection \cite{leo97} and is given by:
\begin{equation}
Q(x_1, x_2, y_1,y_2) = \frac{1}{\pi^2} \langle \alpha_A, \alpha_B |\rho|\alpha_A, \alpha_B \rangle,
\end{equation}
where $|\alpha_{A (B)}\rangle$ is a coherent state centered on $\alpha_{A(B)} = x_{A (B)} + i p_{A (B)}$ in Alice's (Bob's) phase space. 
Interestingly, if we denote $W_0$ the Wigner function of the vacuum, then $Q$ simply corresponds to the convolution of $W$ and $W_0$: $Q= W \star W_0$.

The relation between the two probability distributions $P$ and $Q$ can also be made explicit: if Alice measures one mode of the two-mode squeezed vacuum with a heterodyne detection and obtains outcomes $x_1$, $x_2$, she projects the second mode on the coherent state $|\gamma (x_1 - i x_2)\rangle$ where the factor $\gamma=\sqrt{2(V-1)/(V+1)}$ depends on the variance $V$ of the two-mode squeezed vacuum \cite{GCW03}\footnote{In the PM scenario, Alice's modulation has a variance $V-1$.}.  This means that we have the one-to-one relation:
\begin{equation}
Q(x_1, x_2, y_1,y_2) = P(\gamma x_1, -\gamma x_2, y_1, y_2).
\end{equation}

We now consider general $n$-mode states.
Because of the correspondence above, applying a random transformation $R\otimes R$ with $R$ drawn with the Haar measure on $K(n)$ on the classical data represented by the distribution $P$ is equivalent to a symmetrization in phase-space corresponding to the application of random conjugate passive linear transformations over the $n$ modes of Alice and Bob. Noting $\mathcal{G}$ the group of passive linear transformations in phase-space, the state obtained after the symmetrization of $n$ copies of the state $\rho$ (i.e., for a collective attack) is:
\begin{equation}
\label{symm_state}
\mathcal{S}(\rho^{\otimes n}) := \int_{U \in \mathcal{G}} (U \otimes U^*) \, \rho^{\otimes n}  \, (U \otimes U^*)^\dagger \mathrm{d}U,
\end{equation}
where $\mathrm{d} U$ is the Haar measure over the group $\mathcal{G}$.

\section{Sketch of the security proof}
The rest of the paper consists in analyzing the state $\mathcal{S}(\rho^{\otimes n})$, and in particular proving that it becomes approximately Gaussian under conditions on the second and fourth moment of $P(x_1,x_2,y_1,y_2)$ which are usually met in practical implementations of a CVQKD protocol. Since the three distributions $W$, $Q$ and $P$, equivalently describe the protocol, we choose here to work with $P$, which is the one directly observable in the practical implementation of the protocol.

Our proof will consist of two steps. First, we show that the distribution $P$ describing the state after the symmetrization tends to an explicit limiting function, where the convergence speed is $O(1/\sqrt{n})$. While one could in principle stop at this point and directly compute the secret key rate that can be extracted from this state, we will focus on the experimentally relevant scenario where the quantum channel behaves approximately as a Gaussian channel. In this case, we can bound the distance between the limiting distribution describing the whole protocol and a Gaussian identically and independently distributed (i.i.d.) function (thus corresponding to a collective Gaussian attack for which the key rate is already known) with an error term of order $O(1/\sqrt{n})$.
In such a practical scenario, one can therefore bound the distance between the actual state and the state corresponding to a Gaussian attack by an arbitrary small quantity. Taking $n$ large enough, the secret key rate of the symmetrized protocol is therefore identical to the secret key rate against collective Gaussian attacks.

\section{Convergence to a limiting distribution}
To simplify the notations, we write $\tilde{P}$ the distribution corresponding to the state $\mathcal{S}(\rho^{\otimes n})$. The following lemma from \cite{LG10b} (proven in Appendix \ref{proof-three-variables} for completeness) shows that this function only depends on 3 variables, instead of $4n$ for the non-symmetrized scenario.
\begin{lemma}
\label{three-variables}
For $n \geq 2$, the symmetrized distribution
\begin{equation}
\tilde{P}({\bf x}, {\bf y}) = \int_{K(n)} P(R{\bf x}, R{\bf y}) \mathrm{d} R,
\end{equation}
where $\mathrm{d}R$ refers to the Haar measure on $K(n)$, only depends on $||{\bf x}||^2, ||{\bf  y}||^2, {\bf x} \cdot {\bf y}$.
\end{lemma}

Let us note $X_i = x_i^2, Y_i = y_i^2, Z_i = x_i y_i$ and $X^n = \sum_{i=1}^n X_i, Y^n = \sum_{i=1}^n Y_i, Z^n = \sum_{i=1}^n Z_i$.
Because $\tilde{P}({\bf x}, {\bf y})$ only depends on $X^n=||{\bf x}||^2$, $Y^n=||{\bf  y}||^2$ and $Z^n={\bf x} \cdot {\bf y}$, it actually corresponds to the probability distribution $P_v$ of the vector $V^n=[X^n, Y^n, Z^n]$: 
\begin{equation}
\tilde{P}({\bf x}, {\bf y})  \mathrm{d}{\bf x} \mathrm{d}{\bf y} = P_v(V^n) \mathrm{d} V^n.
\end{equation}

According to the central limit theorem, since the vectors $V_i$ are i.i.d. (which follows from the collective attack assumption), the distribution $P_v$ converges to a Gaussian distribution as $n$ tends to infinity, and one can use a multidimensional version of Berry-Esseen theorem to bound the distance between the two distributions. Noting $P_G$ the Gaussian distribution with the same first two moments as $P_v$, one can prove that the variational distance $\Delta$ between the two distributions is of the form (see Appendix \ref{BE} for a more precise bound):
\begin{equation}
\Delta := \iiint_{\mathbb{R}^3} \left|P_v(V)-P_G(V)) \right| \mathrm{d}V= O\left(\frac{1}{\sqrt{n}}\right).
\end{equation}
The scaling law in $O(1/\sqrt{n})$ is generic for Berry-Esseen theorem and the constant factor depends on the covariance matrix of $[X,Y,Z]$, that is, on the moment of order 4 of the measurement results $(x,y)$.

In general, one could compute the secret key rate corresponding to a state described by the distribution $P_G$. Here, we will restrict ourselves to a very concrete scenario, that of a quantum channel acting like a Gaussian channel (like an optical fiber typically \footnote{Other experimental setups can be considered in practice, for instance free-space CVQKD \cite{HEB10}. In that case, it would be interesting to see whether the quantum channel behaves approximately like a Gaussian channel.}). We insist that this is not a new assumption since one can always compute the covariance matrix $\Sigma$ of the distribution $P_G$ above. However, in general, the quantum channel will be such that $\Sigma$ will be close (up to sampling errors) to the covariance matrix obtained for a Gaussian channel.

\section{The limiting distribution $P_G$ is close to an i.i.d. Gaussian distribution in typical implementations}
In practice, the coherent states are sent through an optical fiber acting as a Gaussian quantum channel, meaning that the data obtained by Alice and Bob follow a Gaussian distribution. In general, this does not simplify the security analysis since observing variables that \emph{look} Gaussian does not mean that they indeed \emph{are} Gaussian. Here, we will show that \emph{looking} Gaussian is sufficient for the bound obtained through Berry-Esseen theorem to be useful. 
Let us consider the case where
\begin{equation}
(x_i, y_i) \sim \mathcal{N}\left(
\begin{bmatrix}
0\\
0\\
\end{bmatrix},
\begin{bmatrix}
\langle x_i^2 \rangle & \langle x_i y_i \rangle \\
\langle x_i y_i \rangle & \langle y_i^2 \rangle \\
\end{bmatrix}
 \right).
 \end{equation}
Then, applying the symmetrization and using the results of Berry-Esseen theorem, one obtains that the symmetrized (normalized) distribution $P_v$ tends to a Gaussian distribution with covariance matrix $\Sigma_G$:
\begin{equation}
\Sigma_G = \left[
\begin{smallmatrix}
3\langle x_i^2 \rangle^2 &  \langle x_i^2 \rangle \langle y_i^2 \rangle + 2 \langle x_i y_i \rangle^2 & 3 \langle x_i^2 \rangle \langle x_i y_i \rangle  \\
\langle x_i^2 \rangle \langle y_i^2 \rangle + 2 \langle x_i y_i \rangle^2 & 3\langle y_i^2 \rangle^2& 3 \langle y_i^2 \rangle \langle x_i y_i \rangle\\
3 \langle x_i^2 \rangle \langle x_i y_i \rangle & 3 \langle y_i^2 \rangle \langle x_i y_i \rangle & \langle x_i^2 y_i^2 \rangle\\
\end{smallmatrix}\right].
\end{equation}
Unfortunately, however, in a practical protocol, Alice and Bob only have access to a finite-precision estimation of the covariance matrix, and the one they measure, $\Sigma_{\mathrm{est}}$, and that they use in Berry-Esseen theorem will slightly differ from the ideal one above $\Sigma_G$. The typical estimation error is of the order of $1/\sqrt{m}$ if $m$ samples are used in the procedure. Assuming that the estimation is performed with a (small) constant fraction of the total samples $n$, the typical error will be on the order of $1/\sqrt{n}$, which is comparable to the error term of the Berry-Esseen theorem. 
This then implies that the variational distance between the two final distributions is also of that order (see Appendix \ref{finite} for details).

\section{Security analysis of a CV QKD protocol with postselection}
Using the results above, the distribution $P$ (or equivalently the $Q$-function) of the state describing the symmetrized version of the QKD protocol is $1/\sqrt{n}$-close in variational distance to a Gaussian distribution. Moreover, in a practical scenario, this Gaussian distribution corresponds to that of an i.i.d. Gaussian state.
If one can make the error $1/\sqrt{n}$ small enough, then the security of the symmetrized protocol against collective attacks is equivalent to that of the usual (non-symmetrized) protocol against Gaussian attacks. 
In particular, the secret key rate for the symmetrized protocol is equal to the secret key rate against Gaussian attacks \cite{HL07,SAA07}.

Although the variational distance between the $Q$-functions is a weaker criterion than the usual trace distance between the states, one can argue that this distance makes sense when considering CV QKD protocols. Indeed, if two states have $2\epsilon$-close $Q$-functions (for the variational distance), it means that the probability to successfully distinguish them using heterodyne detection is bounded by $1/2+\epsilon$.

Sampling from the Haar measure on $K(n)$ for $n$ large might be quite unpractical. Methods to achieve it in complexity $O(n^2)$ are known \cite{LG11, JKL11}. Fortunately, for our purpose, it is sufficient to sample from the different measure on $K(n)$ provided that the symmetrized state can be made arbitrary close to the state $\mathcal{S}\left( \rho^{\otimes n}\right)$ of Eq. \ref{symm_state}. This can be achieved by means of quantum $k$-designs as presented in Appendix \ref{design}. In particular, it is reasonable to conjecture that this can be done in complexity $O(n \log n)$, which would be compatible with a practical implementation. 

One could wonder why the symmetrization has to be \emph{active} here in contrast to protocols such as GG02. The difference is that the postselection, performed along the quadrature axes, introduces some privileged directions in phase space. Consequently one needs to actively symmetrize the protocol to make it invariant under rotations in phase space. 

\section{Conclusion} 
In this paper, we provided a first step towards a general security proof of CVQKD protocols with postselection. Until now, its security was only established in the very restricted case of Gaussian attacks, which are very unlikely to be optimal. Thanks to an active symmetrization of the protocol (performed on the classical data of Alice and Bob), one can show that collective attacks and actually arbitrary attacks in the asymptotic limit basically reduce to Gaussian attacks. 
The present solution is still not very practical since one would need to randomly sample from the unitary group in a very large dimension. Two possible approaches should be considered: either looking at a much smaller set of transformations for which the sampling can be performed efficiently, or improving the bounds derived here, possibly combining the symmetrization technique with some de Finetti-type arguments. 
It seems almost clear, at any rate, that the symmetrization technique introduced here will be required for any further advance in the study of the security of CV QKD with postselection.

\section{Acknowledgments} I thank Antonio Ac\'in, Fr\'ed\'eric Grosshans and Philippe Grangier for fruitful discussions. 
I acknowledge support from the European Union under the ERC Starting grant PERCENT.

\appendix

\section{Proof of Lemma $[1]$ (Lemma [1] from \cite{LG10b})}
\label{proof-three-variables}
Since the probability distribution $P$ is being randomized under the action of the group $K(n) = O(2n,\mathbb{R}) \cap Sp(2n,\mathbb{R})$ to give $\tilde{P}$ defined as \begin{equation}
\tilde{P}({\bf x}, {\bf y}) = \int_{K(n)} P(R{\bf x}, R{\bf y}) \mathrm{d} R 
\end{equation}
where $\mathrm{d}R$ refers to the Haar measure on $K(n)$, one has:
\begin{equation}
\tilde{P}(R{\bf x}, R{\bf y}) = \tilde{P}({\bf x}, {\bf y})
\end{equation}
for any ${\bf x}, {\bf y} \in \mathbb{R}^{2n}$ and $R \in K(n)$.

We want to show that any function $\tilde{P}: \mathbb{R}^{2n} \times \mathbb{R}^{2n}$, such that $\tilde{P}(R{\bf x}, R{\bf y}) = \tilde{P}({\bf x}, {\bf y})$ for any transformation $R \in K(n)$ only depends on the three following parameters: $||{\bf x}||^2, ||{\bf  y}||^2, {\bf x} \cdot {\bf y}$.

Given any four vectors ${\bf x}, {\bf x'}, {\bf y}, {\bf y'} \in \mathbb{R}^{2n}$ such that $||{\bf x}||^2=||{\bf x'}||^2, ||{\bf  y}||^2=||{\bf  y'}||^2, {\bf x} \cdot {\bf y}={\bf x'} \cdot {\bf y'}$, it is sufficient to exhibit a transformation $R\in K(n)$ such that ${\bf x'}=R {\bf x}$ and ${\bf y'}= R {\bf y}$ to prove Lemma $[1]$.

A transformation $R \in K(n)$ can be described as a symplectic map:
\begin{equation}
\label{symplectic}
R = R(X,Y) \equiv
\begin{bmatrix}
X & Y  \\ 
-Y & X  \\ 
\end{bmatrix} 
\end{equation}
where the matrices $X$ and $Y$ are such that \cite{ADMS95}:
\begin{eqnarray}
X^T X+Y^T Y &=& X X^T + Y Y^T =1\\
X^T Y &,& X Y^T \quad \mathrm{symmetric}.
\end{eqnarray}
Note that this matrix is written for reordered vectors of the form $[x_1, x_3, \cdots, x_{2n-1},x_2, x_4, \cdots, x_{2n}]$, that is, one first writes the $n$ $q$-quadratures then the $n$ $p$-quadratures for all the vectors. 

Let us introduce the following vectors ${\bf a}, {\bf a'}, {\bf b}, {\bf b'} \in \mathbb{C}^n$ defined as
\begin{eqnarray}
a_k = x_{2k-1} + i x_{2k} &,& a'_k = x'_{2k-1}+ i x'_{2k}\\
b_k = y_{2k-1} + i y_{2k} &,& b'_k = y'_{2k-1}+ i y'_{2k}.
\end{eqnarray}
Then, the conditions read:
\begin{equation}
\label{condition2}
\left\{
\begin{array}{ccc}
||{\bf a}||^2=||{\bf a'}||^2\\
||{\bf b}||^2=||{\bf b'}||^2\\
\mathrm{Re}\langle {\bf a} | {\bf b} \rangle = \mathrm{Re}\langle {\bf a'} | {\bf b'} \rangle
\end{array}
\right. ,
\end{equation}
where $\mathrm{Re}(x)$ refers to the real part of $x$.

For our purpose, it is therefore sufficient to prove that there exists an unitary transformation $U \in U(n)$ such that $U {\bf a} ={\bf a'}$ and $U {\bf b} = {\bf b'}$. Indeed, one can split $U$ into real and imaginary parts: $U = X - iY$, and it is easy to check that $R=R(X,Y)$ is such that $R{\bf x} ={\bf x'}$ and $R{\bf y} = {\bf y'}$.

Let us introduce the following notations: $A \equiv ||{\bf a}||^2=||{\bf a'}||^2, B \equiv ||{\bf b}||^2=||{\bf b'}||^2$ and $C \equiv \mathrm{Re}\langle {\bf a}| {\bf b} \rangle = \mathrm{Re}\langle {\bf a'}| {\bf b'} \rangle$.

Consider first the case where ${\bf a}$ and ${\bf b}$ are colinear. This means that ${\bf b} = C/A {\bf a}$ and $C = \pm \sqrt{AB}$. 
Using the Cauchy-Schwarz inequality, $|C| = |{\bf a'} \cdot {\bf b'}| \leq ||{\bf a'}|| \cdot||{\bf b'}|| =  \sqrt{AB}$ with equality if and only if ${\bf a'}$ and ${\bf b'}$ are colinear. This means that ${\bf a'}$ and ${\bf b'}$ are colinear and that ${\bf b'} = (C/A) \, {\bf a'}$. 
Because $||{\bf a}|| = ||{\bf a'}||$, the reflexion $U$ across the mediator hyperplane of ${\bf a}$ and ${\bf a'}$ is a unitary transformation that maps ${\bf a}$ to ${\bf a'}$. This reflexion also maps ${\bf b}$ to ${\bf b'}$. This ends the proof in the case where ${\bf a}$ and ${\bf b}$ are colinear.

Let us now consider the general case where ${\bf a}$ and ${\bf b}$ are not colinear. It is clear that ${\bf a'}$ and ${\bf b'}$ cannot be colinear either. We take two bases $({\bf a}, {\bf b}, {\bf f_3}, \cdots, {\bf f_n})$ and $({\bf a'}, {\bf b'}, {\bf f_3'}, \cdots, {\bf f_n'})$ of $\mathbbm{C}^n$ and use the Gram-Schmidt process to obtain two orthonormal bases $\mathcal{B}=({\bf e_1}, \cdots, {\bf e_n})$ and $\mathcal{B}'=({\bf e_1'}, \cdots, {\bf e_n'})$. Note that vectors ${\bf e_1}, {\bf e_2}, {\bf e_1'}$ and $e{\bf _2'}$ are given by:
\begin{eqnarray}
{\bf e_1} = \frac{{\bf a}}{\sqrt{A}} &,& {\bf e_2} = \frac{{\bf b} - \langle {\bf e_1}|{\bf b}\rangle {\bf e_1}}{||{\bf b} - \langle {\bf e_1}|{\bf b}\rangle {\bf e_1} ||}\\
{\bf e_1'} = \frac{{\bf a'}}{\sqrt{A}} &,& {\bf e_2'} = \frac{{\bf b'} - \langle {\bf e_1'}|{\bf b'}\rangle {\bf e_1'}}{||{\bf b'} - \langle {\bf e_1'}|{\bf b}\rangle {\bf e_1'} ||}.
\end{eqnarray}
Let us call $U$ the unitary operator mapping $\mathcal{B}$ to $\mathcal{B}'$. It is easy to see that $U$ maps ${\bf a}$ and ${\bf b}$ to ${\bf a'}$ and ${\bf b'}$, respectively. This concludes our proof.

\section{Distance between the symmetrized distribution and a Gaussian distribution: Berry-Esseen theorem}
\label{BE}
We use a multidimensional local version of the Berry-Esseen theorem due to Zitikis. More precisely, Theorem 1.2 of \cite{Zit93} reads:
\begin{theo}
\label{berry-esseen}
Let $V_1, \cdots, V_n$ be a sequence of independent and identically distributed $d$-variate random vectors, let $S_n = \tfrac1n \sum_{i=1}^n V_i$. Let ${\bf \mu}$ and $\Sigma$ be the first two moments of $V_1$, and let $\lambda_{\mathrm{min}}$ be the least eigenvalue of $\Sigma$. Let $G$ be a Gaussian random vector with zero mean and covariance matrix $\mathbbm{1}_d$. 
Let $\mathcal{C}$ be the class of all convex Borel sets. Then there exists a universal constant $c\geq 0$ such that 
\begin{multline} 
\sup_{C \in \mathcal{C}} \left|P(\sqrt{n} (S_n-\mu)\Sigma^{-1/2}\in \mathcal{C}) - P(G \in \mathcal{C}) \right|\\
\leq c \sqrt{d} \lambda_{\mathrm{min}}^{-3/2}\mathbb{E}\left(||V_1||^3\right)/\sqrt{n},
\end{multline}
where $L_n(1)=1/n \sum_i V_i$, $V_i$ is a $d$-dimensional vector and $||.||$ refers to the Euclidean norm.
\end{theo}
First, we note that the quantity 
\begin{equation}
\sup_{C \in \mathcal{C}} \left|P({\bf x} \in \mathcal{C}) - P({\bf y} \in \mathcal{C})  \right|
\end{equation}
corresponds to the variational distance between the distributions of ${\bf x}$ and ${\bf y}$.

In the case of CVQKD, we need to consider tridimensional random vectors $V_i=[X_i,Y_i,Z_i]$ and one can immediately estimate ${\bf \mu}$ and $\Sigma$ from the experimental data. 
Using the notations of the main text and applying Theorem \ref{berry-esseen} gives
\begin{multline}
\label{bound}
\iiint_{\mathbb{R}^3} \left|P_v(V)-P_G(V)) \right| \mathrm{d}V\\
\leq c \sqrt{3} \lambda_{\mathrm{min}}^{-3/2}\mathbb{E}\left(||X_1^2+Y_1^2+Z_1^2||^{3/2}\right)/\sqrt{n},
\end{multline}
where $P_G$ corresponds to the multivariate Gaussian distribution with the same first two moments as $P_v$.

\section{Error due to the finite estimation sample}
\label{finite}
Let us assume that the variables $(x_k,y_k)$ follow a centered bivariate normal distribution, as typical in experimental implementations of CVQKD.

During the parameter estimation, Alice and Bob need to estimate the fourth moments of $(x_k,y_k)$ given by the covariance matrix $\Sigma$:
\begin{equation*}
\Sigma = \left[
\begin{smallmatrix}
\langle x_i^4 \rangle &  \langle x_i^2 y_i^2 \rangle & \langle x_i^3 y_i \rangle  \\
\langle x_i^2 y_i^2 \rangle & \langle y_i^4 \rangle & \langle x_i y_i^3 \rangle\\
\langle x_i^3 y_i \rangle & \langle x_i y_i^3 \rangle & \langle x_i^2 y_i^2 \rangle\\
\end{smallmatrix}\right]
\end{equation*}
Alternatively, one can describe this matrix by the vector $V_i = [\langle x_i^4 \rangle, \langle x_i^3 y_i \rangle, \langle x_i^2 y_i^2 \rangle, \langle x_i y_i^3 \rangle, \langle y_i^4 \rangle]$.
In order to estimate this vector, one uses the estimator $\bar{V}^m$ defined as
\begin{equation}
\bar{V}^m:= \frac{1}{\sqrt{m}} \sum_{i_1,\cdots, i_m} V_i,
\end{equation}
where $i_1,\cdots, i_m$ are $m$ randomly chosen indices among $\{1,\cdots,n\}$.
Using the mutlivariate version of the Central Limit Theorem, one obtains that the estimator $\bar{V}^m$ converges to the true value and that the error follows a normal distribution.

More precisely, let us note $\Sigma_G$ the true covariance matrix and $\Sigma_{\mathrm{est}}$ the estimated matrix, i.e.,
\begin{equation}
\Sigma_G = \left[
\begin{smallmatrix}
3\langle x_i^2 \rangle^2 &  \langle x_i^2 \rangle \langle y_i^2 \rangle + 2 \langle x_i y_i \rangle^2 & 3 \langle x_i^2 \rangle \langle x_i y_i \rangle  \\
\langle x_i^2 \rangle \langle y_i^2 \rangle + 2 \langle x_i y_i \rangle^2 & 3\langle y_i^2 \rangle^2& 3 \langle y_i^2 \rangle \langle x_i y_i \rangle\\
3 \langle x_i^2 \rangle \langle x_i y_i \rangle & 3 \langle y_i^2 \rangle \langle x_i y_i \rangle & \langle x_i^2 y_i^2 \rangle\\
\end{smallmatrix}\right],
\end{equation}
and
\begin{equation}
\Sigma_\mathrm{est} = \frac{1}{m} \sum_{i_1, \cdots, i_m}\left[
\begin{smallmatrix}
x_i^4 & x_i^2 y_i^2 & x_i^3 y_i \\
x_i^2 y_i^2 &  y_i^4  &  x_i y_i^3 \\
x_i^3 y_i  & x_i y_i^3  &  x_i^2 y_i^2\\
\end{smallmatrix}\right].
\end{equation}
Then, the Central Limit Theorem asserts that the random matrix $\sqrt{m} (\Sigma_\mathrm{est}-\Sigma_G)$ converges in distribution to a centered multivariate normal distribution with a covariance matrix depending on moments of order 8 of $(x_k,y_k)$. We do not explicitate this matrix here as it is rather cumbersome, but it is straightforward to compute it.

The result of this analysis is that the error in estimating the correct covariance  matrix $\Sigma_g$ scales as $1/\sqrt{m}$ where $m$ is the number of samples used. In order to obtain an error of the same order of magnitude as the one due to Berry-Esseen theorem, one should use a constant fraction of the data for the parameter estimation. In that case, the covariance matrix would be estimated with a precision $1/\sqrt{n}$.

We now prove that this error for the covariance matrices translates into an error (computed with respect to the variational distance) of the same magnitude for the probability distributions.

Let us first consider the case of univariate normal distributions, that is two normal distributions $\mathcal{N}(0,\sigma_1^2)$ and $\mathcal{N}(0,\sigma_2^2)$ with $\sigma_2 > \sigma_1$. Let us note $g_1$ and $g_2$ their respective density function.
One has:
\begin{eqnarray}
&& \int_{-\infty}^{\infty} |g_1(x) - g_2(x)| \mathrm{d}x \nonumber \\
&&= 2 \mathrm{erf} \left(\sigma_2 \sqrt{\frac{\ln (\sigma_2/\sigma_1)}{\sigma_2^2-\sigma_1^2}} \right)-2 \mathrm{erf} \left(\sigma_1 \sqrt{\frac{\ln (\sigma_2/\sigma_1)}{\sigma_2^2-\sigma_1^2}} \right)  \nonumber
\end{eqnarray}
where $\mathrm{erf}(x) = 2/\sqrt{\pi} \int_0^x e^{-t^2}\mathrm{d}t$ is the error function.
In particular, if one has $\sigma_2 = \sigma_1 + \delta$ where $\delta=O(1/\sqrt{n})$ is a small error, then a first order expansion of the expression above gives
\begin{equation}
\int_{-\infty}^{\infty} |g_1(x) - g_2(x)| \mathrm{d}x = \delta \sqrt{\frac{8}{e \pi \sigma_1^2}} + o\left(\frac{1}{\sqrt{n}}\right).
\end{equation}

One can extend this analysis to the case of multivariate normal distributions with covariance matrices differing by an error of the order of $1/\sqrt{n}$, and one would get that the variational distance between the two distributions is also of the order of $1/\sqrt{n}$.
For lisibility, we omit the precise bound here but it can be computed in a straightforward manner.

\section{Approximate symmetrization using efficient construction of quantum $k$-designs}
\label{design}
Sampling from the group $K(n)$ is equivalent to sampling from the unitary group $U(n)$. Indeed, any map in $K(n)$ can be described by its action on the annihilation operators on the $n$ modes. Let us note $a_i$ (resp. $b_i$) the annihilation operator on the $i^\mathrm{th}$ mode of Alice (resp. Bob). 
A map $U \in K(n)$ is described by the unitary $u_{i,j}$ that transforms $a_i$ into $\sum_{j=1}^n u_{i,j} a_j$.
Let us consider a general state $\rho \in \left(\mathcal{H}_A\otimes \mathcal{H}_B\right)^{\otimes n}$,
\begin{equation}
\rho = \sum_{{\bf i}^a, {\bf i}^b, {\bf j}^a,{\bf j}^b} \lambda_{{\bf i}^a{\bf i}^b {\bf j}^a {\bf j}^b} |{\bf i}^a, {\bf i}^b\rangle \!\langle {\bf j}^a, {\bf j}^b|,
\end{equation}
where ${\bf i}^a = (i_1^a, i_2^a, \cdots, i_n^a)$ describes the photon distribution in Alice's $n$ modes (and similarly for Bob with ${\bf i}^b$).

The state $ |{\bf i}^a, {\bf i}^b\rangle = \frac{1}{\sqrt{{\bf i}^a! {\bf i}^b !}} \prod_{k=1^n} \left( a_k^{\dagger}\right)^{i_k^a} \left( b_k^{\dagger}\right)^{i_k^b} |00\rangle$ is transformed under the action of $U\otimes U^*$ into
\begin{multline}
U\otimes U^* |{\bf i}^a, {\bf i}^b\rangle= \\
\frac{1}{\sqrt{{\bf i}^a! {\bf i}^b !}} \prod_{k=1^n} \left( \sum_{l=1}^n u_{k,l} a_l^{\dagger}\right)^{i_k^a} \left(\sum_{l=1}^n u_{k,l}^* b_l^{\dagger}\right)^{i_k^b} |00\rangle.
\end{multline}

Let us fix a maximal photon number $N$ together with the projector $\Pi_{N}$ on the subspace of $\mathcal{H}_A\otimes \mathcal{H}_B$ spanned by Fock states $ |{\bf i}^a, {\bf i}^b\rangle$ with a total photon number less or equal to $N$.
For some given first four moments of $\rho$, there exists a constant $c_\epsilon$ such that taking $N=  c_\epsilon n$ leads to 
\begin{equation}
\left\| \rho - \Pi_{N} \rho  \Pi_{N}^\dagger \right\|_\mathrm{tr} \leq \epsilon.
\end{equation}
Noting $\rho_{N} =\Pi_{N} \rho  \Pi_{N}^\dagger$, one observes that $(U \otimes U^*)  \rho_{N}  (U \otimes U^*)^\dagger$ is a polynomial of degree $N$ in $u_{i,j}$. 
Consider an approximate $N$-design $\nu$, then for any $k \leq N$, one has
\begin{multline}
\left\| \int_{\mathrm{Haar}} (U \otimes U^*) \, \rho^{\otimes n}  \, (U \otimes U^*)^\dagger \mathrm{d}U \right. \\
\left.  -   \sum_\nu (U(\nu) \otimes U^*(\nu) ) \, \rho^{\otimes n}  \, (U(\nu)  \otimes U^*(\nu) )^\dagger \right\|_\mathrm{tr} \leq \epsilon_\mathrm{design}.
\end{multline}
Hence, it is sufficient to use an approximate  $N$-design instead of the Haar measure on the unitary group $U(n)$ in order to symmetrize the state $\rho_{N}$, up to some arbitrary small error $\epsilon_\mathrm{design}$.
Interestingly, efficient constructions of such approximate $N$-designs are known \cite{HL09}. In these constructions, the number of unitaries in the design scales as $n^{O(N)}$ which means that one needs $O( n \log n)$ bits of randomness to draw one such unitary (since $N$ is proportional to the number of modes). 
Unfortunately, the construction provided in \cite{HL09} only works in the regime where $N= O(n/ \log n)$, which is close to, but not exactly the regime of interest here. 
Nevertheless, the existence of this construction gives hope that one could construct an approximate $N$-design also in the regime where $N = O(n)$. If this were true, then one could efficiently (in time $O(n \log n)$) perform the symmetrization studied in the main text.

\end{document}